# Atomic Layer Etching of Aluminum Nitride: Mechanistic Insights from First-Principles Studies of Chlorine Chemistry

Sanjay Nayak[1*], Nikolai Andrianov[2], Thomas Gruhn[3], and Joaquín Miranda[1]

[1]Thin Film Research Unit, Microsystem Division, Silicon Austria Labs GmbH, Villach 9524, Austria

[2]Microfab Division, Silicon Austria Labs GmbH, Villach 9524, Austria

[3] Department of Biomaterials, Faculty of Engineering Science, University of Bayreuth, Prof.-Rüdiger-BormannStr. 1, Bayreuth 95447, Germany

Using *first-principles* density functional theory (DFT) calculations in combination with the climbing-image nudged elastic band (CI-NEB) method, we investigated the adsorption, desorption, and diffusion of atomic chlorine (Cl) and molecular chlorine ($Cl_2$) on the Al-terminated (0001) surface of aluminum nitride (AlN). Our results reveal that both atomic Cl and $Cl_2$ exhibit a chemisorption character with high binding energies. Calculations revealed that the splitting pathway of $Cl_2$ on the Al-terminated AlN(0001) surface is a barrierless and exothermic process. These findings provide new microscopic-scale insights into halogen–insulator surface interactions and opportunities for new strategies in optimizing AlN atomic layer etching process in semiconductor fabrication.

*Corresponding Author: sanjay.nayak@silicon-austria.com,
sanjaynayak.physu@gmail.com (SN)

The contemporary nanoscale and sub-nanoscale semiconductor device manufacturing demands atomically precise manufacturing techniques[1–3]. As functional device size continues to shrink and complexity increases, the demand for gentle material etching processes that produce atomically smooth surfaces has increases[4]. Atomic layer etching (ALE) technique that remove materials with atomic precision while preserving the intrinsic properties by avoiding the bulk defect formation unlike conventional continuous etching[4] is at the forefront of modern nanoelectronics device fabrication. Plasma-enhanced atomic layer etching (PEALE) has demonstrated remarkable success in achieving the surface quality required for advanced CMOS applications[5–7]. In last few years many studies are carried out to understand the atomic scale chemistry and physics of the PEALE process. Special attention is given to Si and $SiO_2$ owing to their critical importance in CMOS technology[8–18].

A standard PEALE process typically involves a cyclical, multi-step process: The basic steps are as follows[19]:

1. **Surface modification phase or Dosing step**: In this process the surface is exposed to reactive etching gas, often dissociated or plasma activated halogens. Once the gas is adsorbed on the surface it chemically modifies the top atomic layers. The typical reactive etching gases are chlorine based ($Cl_2$, $BCl_3$), and fluorine based ($SF_6$, $CF_4$, $C_4F_8$).
2. **Purge step**: Post surface modification the reaction chamber is purged with inert gas, e.g. Ar, to remove the unreacted reaction byproducts from the surface.

3. **Removal step (or Ion bombardment step)**: In this step, the surface is bombarded with inert gas ions such as Ar⁺ with an optimal kinetic energy, also called ad threshold energy. The ion bombardment transfers energy to the modified surface layer, facilitating its conversion into volatile or unstable byproducts that desorb and leave the surface, effectively removing one atomic layer without damaging the underlying material.
4. **Second purge step**: Another purge step that removes the reaction products and residuals before the cycle repeats until the desired thickness is achieved.

Following the successful adoption of ALE for advanced CMOS technology development, its utility has been increasingly explored across development of other semiconductor technologies[20,21]. ALE's atomic-scale precision and minimal underlying surface damage make it especially suitable for etching complex engineered 3D nanostructure for next generation logic devices, such as gate-all-around transistors[22,23] and advanced memory structures[24]. Additionally, ALE is being seen as de-facto etching method for compound semiconductors like group III-V semiconductors and their derived technologies [20,25]. Amongst them AlN and its ternary alloy aluminum gallium nitride ($Al_{1-x}Ga_xN$) are extensively employed in micro- and nanoelectronics systems, particularly in photonic structures-such as waveguides, metasurfaces, high electron mobility transistors (HEMTs) and MEMS devices encompassing acoustic, optoelectronic, and the sensing components[26]. This is largely due to high thermal conductivity, wide bandgap, and strong piezoelectricity of AlN[27,28]. A key challenge in fabricating AlN- and AlGaN-based devices lies in achieving nanoscale etching precision while maintaining low surface roughness and the vertical etch profiles. In photonics, deviations from the vertical sidewalls of photonics pillar can degrade the waveguide confinement and coupling efficiency, induce birefringence, and introduce phase errors in the metasurfaces[29,30]. Defective surface in the gate recess etching and surface treatments in AlN- and AlGaN-based HEMTs limits the device performance [21]. Thus, ALE is seen an essential technique in the fabrication process for next generation high performance photonics and HEMT technologies.

While the chemistry of PEALE process for material etching for Si and $SiO_2$ is well established, the corresponding surface reactions for III-V semiconductors including group III-nitrides (AlN, GaN, InN, and their ternary alloys) remain largely unexplored. In this work, we employ *first-principles* DFT simulations to investigate the adsorption, desorption, and diffusion of atomic Cl and $Cl_2$ on the polar Al-terminated (0001) surface of AlN. This polar surface is important as the most HEMT and photonics devices are fabricated on it. The polar discontinuity induced 2DEG at the interface of GaN/AlGaN layer is also formed when the films are grown along the [0001] crystallographic direction.

The rate of adsorption ($r_{st}^{ad}$) and desorption ($r_{st}^{des}$) of any adsorbate on the specific surface atomic site (st) of a solid surface is given by[31];

$$r_{st}^{ad}(T, p_i) = f_{st,i}^{ad}(T) \times \frac{A_{st,i}}{A_{uc}} \times \exp\left(-\frac{\Delta E_{st,i}^{ad}}{k_B T}\right) \times \frac{p_i A_{uc}}{\sqrt{2\pi m_i K_B T}} \quad (1)$$

$$r_{st}^{des}(T, p_i) = f_{st,i}^{des}(T) \times \frac{A_{st,i}}{A_{uc}} \times \frac{K_B T}{h} \times \exp\left(-\frac{\Delta E_{st,i}^{des}}{k_B T}\right) \quad (2)$$

Where the factors $f_{st,i}^{ad}(T)$ and $f_{st,i}^{des}(T)$ are account the sticking probability and are often less than 1. The factors $\frac{A_{st,i}}{A_{uc}} \times \exp\left(-\frac{\Delta E_{st,i}^{ad}}{k_B T}\right)$ represent the local sticking coefficient, where $\Delta E_{st,i}^{ad}$ is the adsorption barrier of adsorbate from a gas phase to transition phase. $A_{uc}$ represent the area of surface unit cell, and $A_{st,i}$ is the area of the adsorption site. $\Delta E_{st,i}^{des}$ represent desorption barrier from condensation state to transition state. $K_B$, $T$, and $p_i$ represent Boltzmann constant, absolute temperature, and partial pressure respectively.

The $\Delta E_{st,i}^{ad}$ and $\Delta E_{st,i}^{des}$ are also related through the binding energy ($E_{b,st,i}$) of the adsorbate to the surface site by:

$$\Delta E_{st,i}^{des} = \Delta E_{st,i}^{ad} - E_{b,st,i} \tag{3}$$

The rate of diffusion of adsorbate from one atomic surface site ($st$) to another surface site ($st'$) is given by:

$$r_{st,st',i}(T) = f_{st,st',i}^{diff,TST}\left(\frac{k_BT}{h}\right)\exp\left(-\frac{\Delta E_{st,st',i}}{k_BT}\right) \tag{4}$$

Where $f_{st,st',i}^{diff,TST} = \frac{q_{TS(st,st',i)}^{vib}}{q_{st,i}^{vib}}$, $q_{TS(st,st',i)}^{vib}$ is partition function of the barrier between $st$ and $st'$ sites, and $q_{st,i}^{vib}$ is the partition function of the initial bound state. $\Delta E_{st,st',i}$ is the maximum barrier for diffusion from $st$ to $st'$ site.

In our study, we determine the binding energy, $E_{b,st,i}$, also known as the adsorption energy ($E_{ads}^{st}$) of Cl and Cl2 adsorbate at different surface sites at low surface coverage of the solid surface and is given by:

$$E_{ads}^{st} = E_{tot.}(\text{surface+adsorbate}) - E_{tot.}(\text{surface}) - E_{tot.}(\text{adsorbate}) \tag{5}$$

where $E_{tot.}$ (surface+adsorbate) is the free energy of the surface and adsorbate present at specific site ($st$) and $E_{tot.}$ (surface) is the total free energy of the pristine surface. The $E_{tot.}$ (adsorbate) are the total free energies of the isolated adsorbate, here they are isolated atomic Cl and $Cl_2$ molecule.

In computing the total free energy of the surfaces, and adsorbates, we used *first-principles* DFT simulation using Quantum Espresso code[32]. The exchange and co-relation functional term is approximated to PBE version of generalized-gradient-approximation (GGA)[33]. Projected augmented wave[34] type pseudopotential is used with a non-linear core correction. The plane wave's kinetic energy cut-off of 610 eV and charge density cutoff of 4350 eV are used in our simulations. The valence electronic configurations of the atoms Al, N, Cl, and H are $3s^2\ 3p^1$, $2s^2\ 2p^3$, $3s^2\ 3p^5$, and $1s^1$ respectively.

The Al-terminated (0001) surface of wurtzite phase AlN (w-AlN) is constructed within the slab model, where a total of 13.2 Å of AlN and a 20 Å of vacuum is used for minimum image interaction along z-direction. The in-plane cell dimensions of the slab are multiple of the $a\ (= b)$ lattice parameter of e bulk w-AlN. The optimal $a$ lattice parameter of bulk w-AlN is obtained by minimizing the total free energy as a function of cell volume. In obtaining the total free energy of bulk w-AlN, the Brillouin zone is sampled with a Γ-centered $k$-point mesh of 8×8×5 [35]. The atomic positions of all ions are relaxed until forces on each atom are less than $10^{-3}$ eV/Å. Adsorption energy ($E_{ads}$) and diffusion of Cl and $Cl_2$ molecules are studied with a supercell slab of $3a×3b$ in-plane dimension. The $k$-points for slab calculations are scaled to appropriate values as optimized from the bulk w-AlN calculation. Neutral hydrogen passivation is used at the bottom part of the slab. The total free energies of the isolated Cl and $Cl_2$ molecules are computed by keeping them in a cubic cell of 10×10×10 Å$^3$ dimension.

The barrier for the diffusion, $\Delta E_{st,st',i}$ of the atomic Cl and the splitting of Cl$_2$ into Cl atoms are computed using climbing image nudge elastic band (CI-NEB) techniques, where the initial and final states are frozen[36].

The optimized lattice parameters of bulk w-AlN are $a$=3.13 Å and $c$=5.01 Å, with an $u$ value of 0.318. The basal plane Al–N bond length ($d_{Al-N}$) is 1.90 Å. The computed GGA-PBE bandgap of w-AlN is 4.05 eV, which underestimates the experimental value, consistent with the limitations of the DFT-GGA[37]. The lattice parameters ($a$ and $c$), agree well with previous reports and are overestimated as commonly reported in the literature that employed a similar methodology as ours[37]. The Cl-Cl bond length ($d_{Cl-Cl}$) in Cl$_2$ molecule is estimated to 2.00 Å, a value consistent with earlier published data[38].

Figure 1a shows the relaxed atomic structure of the pristine (0001) slab. The basal plane $d_{Al-N}$ of top-surface layer is 1.927 Å, 1.42% longer compared to that of bulk w-AlN. The elongation in the bond is due to the relaxation of surface Al atom outward to the vacuum. Figure 1b represents a constant-height simulated scanning tunnelling Micrograph of the surface, with symmetric atomic adsorption sites are indicated. Figures 1c and 1d display the electronic band structure and the density of states (DOS) of the surface slab, respectively. The Fermi level (E$_F$) lies within the conduction band (Figure 1c-d), indicating occupied surface states located below the bulk conduction band. These surface states are dominated by Al-2p orbitals, followed by Al-2s orbitals, as shown in Figure 1d. Our analysis also show the subsurface N-2p states also contribute towards the surface states at the bottom of the bulk conduction band edge.

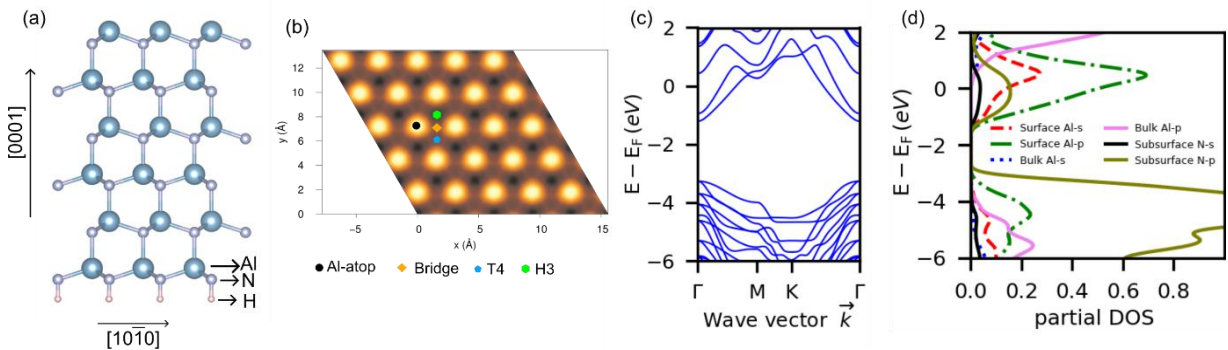

Figure 1: (a) Ball and stick model of the slab considered in our simulation. (b) Simulated scanning tunnelling microscopy images of the Al terminated 0001 surface of AlN. Different atomic adsorption sites are marked for reference. (c) and (d) are the electronic band structure of the slab and the atom/orbital resolved density of states from the surface and bulk Al atoms respectively.

The adsorption energies of atomic Cl ($E_{ads}^{Cl}$) on different symmetric surface adsorption sites of the Al-terminated AlN(0001) surface are summarized in Table 1. The results indicate that Cl adsorption is chemisorption in nature. The most favourable adsorption site is the Al-bridge site, with an $E_{ads}^{Cl}$ of –5.53 eV/atom. We estimate the desorption energies for each symmetric site by calculating the total energy of the slab-plus-adsorbate system as a function of the slab's surface and the vertical Cl atom position (z-distance) (see Figure 2a). The results show that desorption requires a very high energy, indicating strong Cl binding to the surface.

For the lowest energy configurations of each symmetric surface adsorption site, the computed Al–Cl bond lengths ($d_{Al-Cl}$) are tabulated in Table 1. The $d_{Al-Cl}$ in the most stable adsorption configuration (i.e., Al-bridge) is 2.44 Å, close to the $d_{Al-Cl}$ of 2.34 Å found in the stable monoclinic AlCl$_3$ compound (space group: C2/m)[39].

Table 1: shows the site dependence of adsorption energy ($E_{ads}^{Cl}$), desorption energy ($\Delta E_{st,i}^{des}$) and optimal bond length of Al-Cl ($d_{Al-Cl}$) in lowest energy configurations at each surface adsorption sites.

| Adsorption Sites | $E_{ads}^{Cl}$ (eV/atom) | $\Delta E_{st,i}^{des}$ (eV/atom) | Optimal $d_{Al-Cl}$ (Å) |
|---|---|---|---|
| T4 | -4.95 | 4.91 | 2.70 |
| H3 | -5.30 | 5.33 | 2.59 |
| Al-atop | -5.41 | 5.51 | 2.15 |
| Bridge | -5.53 | 5.55 | 2.44 |

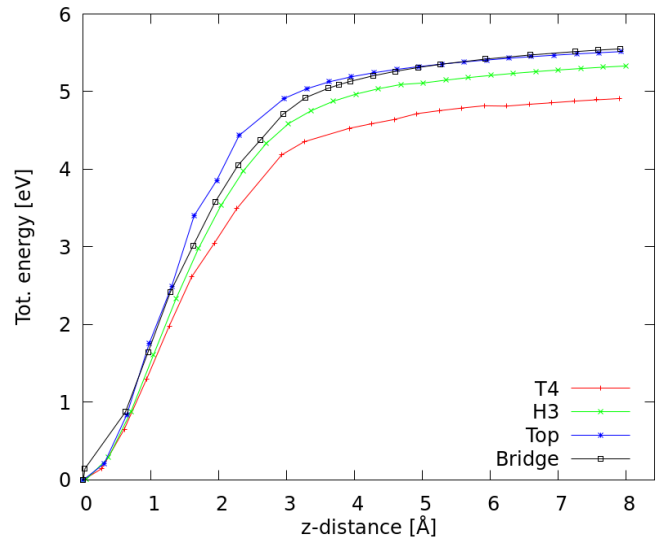

Figure 2: Shows the adsorption energies of Cl on AlN(0001) surface as a function of z-distance i.e. the separation between the adsorbed site at the surface and the Cl atom position relative to vertical position to this site.

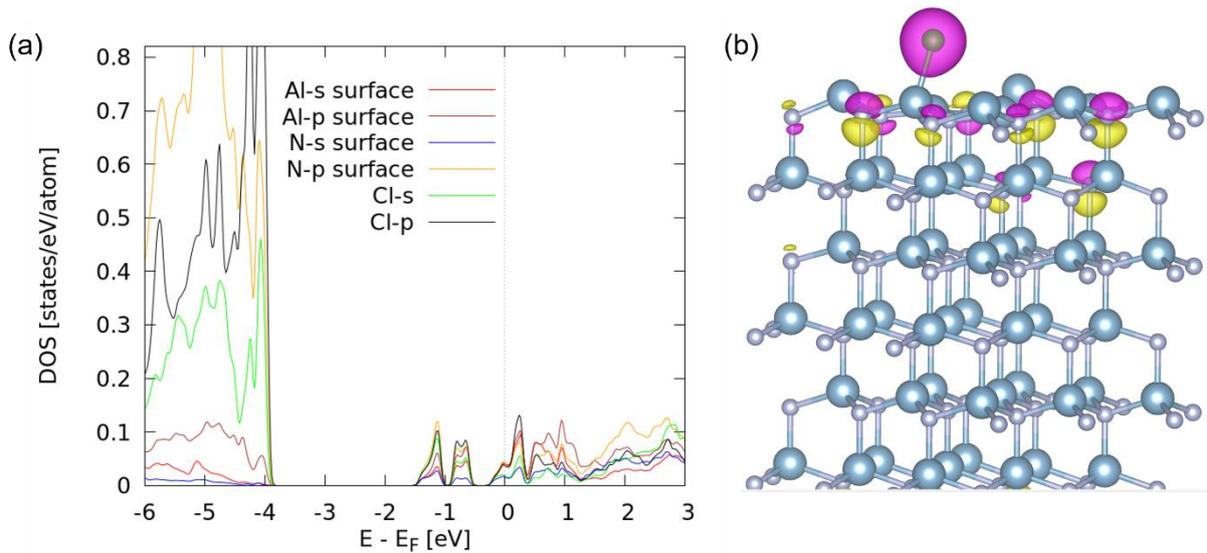

Figure 3: (a) Electronic density of states of Cl adsorbed at a bridge site of Al-terminated AlN(0001) surface. (b) Shows charge density difference ($\rho_{diff} = \rho_{slab+Cl} - \rho_{slab}$).

Analysis of the electronic DOS of the Al-bridge site Cl adsorption revealed the formation of two distinct electronic sub-bands within the bulk bandgap, as illustrated in Figure 3a. Both sub-bands are fully occupied. Atom and orbital projected DOS analysis shows that the largest contribution to these sub-bands originates from the subsurface N atoms, specifically the N-2p orbitals. The orbital contributions are dominated by Cl-3p, with lesser but still significant contributions from Cl-3s, Al-2p, and Al-2s. Notably, the lower-energy sub-band exhibits a greater contribution from Cl-3s states compared to the sub-band with higher energy. Examination of the surface electronic structure shows that the valence band edge is predominantly dominated by the Cl-3p orbitals, with subsequent contributions from the N-2p and Cl-3s orbitals. The charge density difference ($\rho_{diff} = \rho_{slab+Cl} - \rho_{slab}$) revealed that the adsorption process induces an electronic modifications that extend up to half the unit cell length, from the surface, along the c-direction.

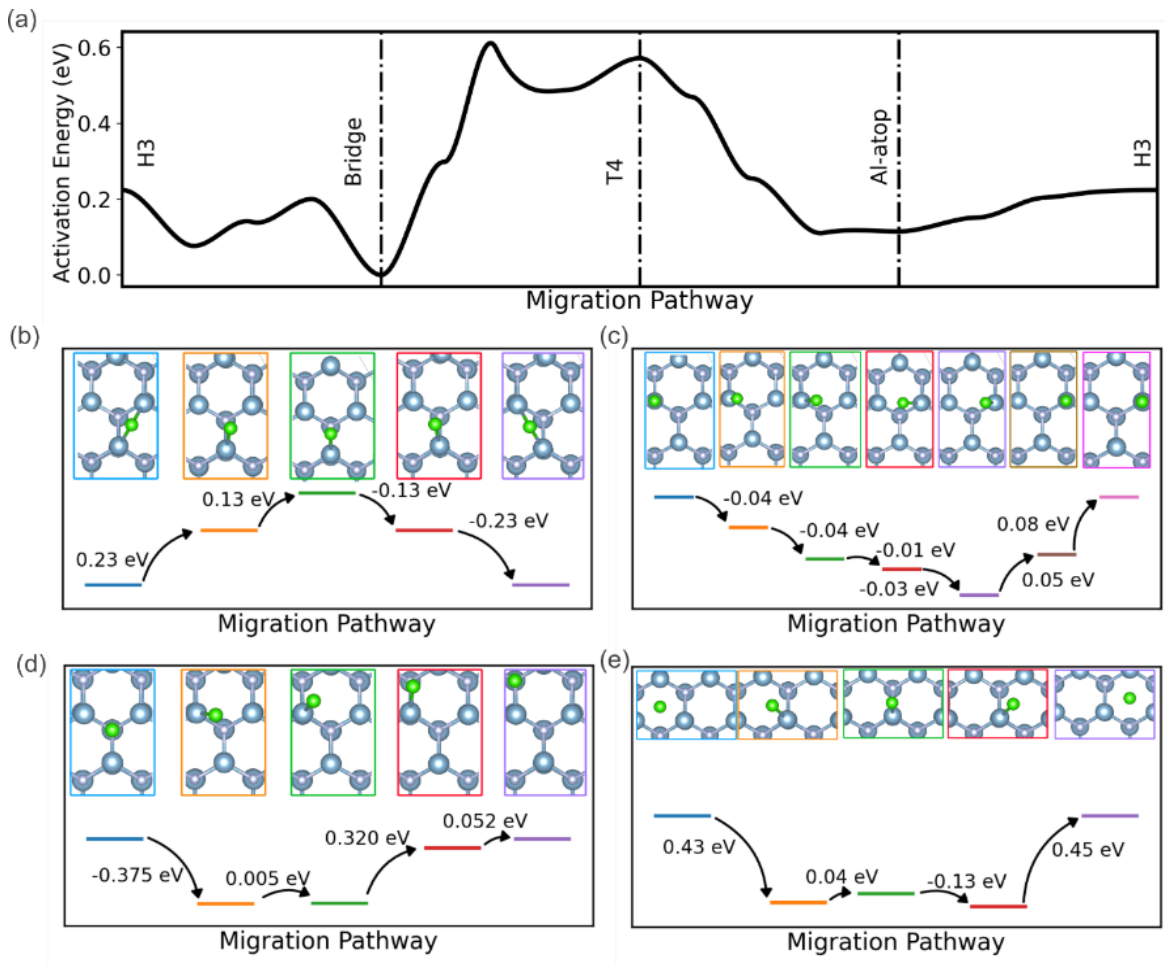

Figure 4:(a) Activation energy, $E_a$, for the diffusion of Cl atoms across the path H3 → Al-bridge → T4 → Al-atop →H3. The energy barrier for the diffusion of Cl atoms from Al bridge → Al-bridge site (a), Al-atop → Al-atop site (c), T4 → T4 site (d) and H3 → H3 site (e) are presented.

Next, we evaluated the activation energies ($E_a$) of atomic Cl for diffusion on Al-terminated AlN (0001) surface. We considered the diffusion of atoms between initial and final sites to be both symmetric and asymmetric (see Figure 4). The $E_a$ of diffusion between asymmetric points are presented in Figure 4a and values are tabulated in Table 2. Our analysis suggests that the $E_a$ is higher when the Cl-adatom diffusion direction is from the three-fold Al atom coordinated surface site (e.g., H3 and T4) to the two-fold co-ordinated surface site and the one-fold co-ordinated surface sites (Bridge and Al-atop). But the $E_a$ for the reverse migration from the two-fold and one-fold co-ordinated surface sites to the three-fold coordinated surface site is lower (see Table 2).

Similarly, the E_a for diffusion from the two-fold co-ordinated surface Al-bridge site to the one-fold co-ordinated surface site (Al-atop) follows similar trend. It is noted that the diffusion of Cl adatoms from the Al-bridge to Al-atop sites can occur through two different paths: (i) in one case the diffusion can occurs within one conventional unit cell via H3 site, (ii)the diffusion can occur to next conventional unit cell and diffused through the three-fold co-ordinated T4 site. Our calculation shows that the E_a for the latter is lower compared to the former ones. We further computed the energy pathways for diffusion of Cl atom between symmetric sites and identified their saddle points (see Figure 4 b-e and Table 2).

Table 2: tablulates the activation energy for diffusion for Cl atoms in both forward and reverse direction.

| Diffusion Path | $E_a$ (eV) | Diffusion Path | $E_a$ (eV) | Diffusion Path | $E_a$ (eV) |
|---|---|---|---|---|---|
| H3 → Bridge | 0.000 | Bridge → H3 | 0.223 | Bridge → Bridge | 0.542 |
| Bridge → T4 | 0.595 | T4 → Bridge | 0.024 | Bridge → Al-atop Bridge | 0.386 |
| T4 → Al-atop | 0.000 | Al-atop → T4 | 0.456 | T4 → Al-atop → T4 | 0.322 |
| Al-atop → H3 | 0.109 | H3 → Al-atop | 0.000 | H3 → H3 | 0.452 |
| Bridge → T4 → Al-atop | 0.524 | Al-atop → T4 → Bridge | 0.410 | Al-atop → Al-atop | 0.132 |
| Bridge → Al-atop | 0.684 | Al-atop → Bridge | 0.000 | - | - |

For the atomic Cl diffusion from one Al-bridge site to an adjacent Al-bridge site, we find that the saddle point is 0.36 eV higher in energy than the initial and final states (see Figure 4b).Diffusion from one Al-atop site to another Al-atop via a Al-bridge site, the lowest energy configuration is neither at the Al-atop or the Al-bridge position. Instead, the energy is at a minimum when the Cl coordinate is between the Al-bridge and Al-atop sites, with an estimated $d_{Al-Cl}$ of 2.18 Å (see Figure 4c). This means that the saddle point for this path is the Al-atop site, which is 0.12–0.13 eV higher in energy than the lowest energy state. We see a similar energy landscape for both the T4-to-T4 and H3-to-H3 diffusion paths. In these cases, the saddle points are approximately 0.375 eV and 0.45 eV above their respective low-energy configurations (see Figure 4d-e).

Further, we computed the adsorption energy of a $Cl_2$ molecule ($E_{ads.}^{Cl_2}$) on the Al-terminated AlN (0001) surface, considering two main orientations:

1. The Cl-Cl bond axis is parallel to the [0001] crystallographic direction (i.e., perpendicular to the (0001) crystal planes).
2. The Cl-Cl bond axis is perpendicular to the [0001] crystallographic direction (i.e., parallel to the (0001) crystal planes).

In both cases the $d_{Cl-Cl}$ is fixed at 2.00 Å. In the former case, we identify four symmetric surface adsorption sites, as we discuss for the atomic Cl case while for the latter case, we consider five distinct initial surface adsorption configurations (see Table 3).

For configurations where the bond axis is parallel to the [0001] direction, the most preferred surface site for adsorption is the Al-bridge site, with $E_{ads.}^{Cl_2}$ of -1.30 eV/molecule. The minimum $d_{Al-Cl}$ of this configuration is 2.60 Å. The other three configurations show similar adsorption energies, ranging from -0.90 to -1.04 eV/molecule, with $d_{Al-Cl}$ varying from 2.30 to 2.90 Å (see Table 3).

For configurations (Conf.5-9 in Table 3) where the $Cl_2$ bond axis is perpendicular to [0001] direction, the lowest $E_{ads.}^{Cl_2}$ (= -2.79 eV/molecule) is with the configuration when one Cl atom is between bridge and H3 surface site while another Cl atom is along the Al-N bond axis (see conf.9 of Table 3). The $d_{Al-Cl}$ in this case is in the range of 2.29-2.47 Å. The second most low energy configuration is the Conf.5 of Table 3, where both Cl atoms are near the Al-atop sites. In this configuration, $d_{Al-Cl}$ is approximately 2.21 Å.

Three other configurations (Conf.6,7, and 8) that are tabulated in Table 3 shows substantially higher $E_{ads.}^{Cl_2}$. The estimated $E_{ads.}^{Cl_2}$ in configurations where the Cl$_2$ bond axis is perpendicular to [0001] are energetically more favourable over the conditions where the Cl-Cl bond axis is parallel to [0001].

Table 3: Calculated adsorption energies of Cl$_2$ molecule, $E_{ads.}^{Cl_2}$, on different sites on the AlN(0001) surface. The lattice vectors of the slab for calculations are a = [9.3957, 0, 0] Å, b= [-4.69785, 8.1369, 0] Å, and c= [0, 0, 33.146] Å. The dark sphere represents the Al site; white sphere represents N site and green sphere represent Cl atoms sites. The optimal Al-Cl bond length, d$_{Al-Cl}$, is also mentioned.

| Configuration | Configuration Name | Atomic Structure | Coordinates of Cl atoms | $E_{ads.}^{Cl_2}$ (eV/molecule) | d$_{Al-Cl}$ (Å) |
|---|---|---|---|---|---|
| Cl-Cl bond axis ∥ [0002] | Conf.1 (Al-atop) | | [0.44444, 0.55555, 0.76216] | -1.04 | 2.30 |
| | | | [0.44444, 0.55555, 0.85900] | | |
| | Conf.2 (Bridge) | | [0.61111, 0.38889, 0.76192] | -1.30 | 2.60 |
| | | | [0.61111, 0.38889, 0.82241] | | |
| | Conf.3 (H3) | | [0.33333, 0.33333, 0.75919] | -1.00 | 2.74 |
| | | | [0.33333, 0.33333, 0.81968] | | |
| | Conf.4 (T4) | | [0.55555, 0.44444, 0.76429] | -0.90 | 2.90 |
| | | | [0.55555, 0.44444, 0.82478] | | |
| Cl-Cl bond axis ⊥ [0002] | Conf.5 | | [0.47781, 0.54425, 0.76294] | -2.77 | 2.21 |
| | | | [0.71168, 0.59495, 0.75965] | | |
| | Conf.6 | | [0.59199, 0.45824, 0.74000] | -1.49 | 2.08-2.40 |
| | | | [0.44216, 0.55389, 0.76000] | | |
| | Conf.7 | | [0.49655, 0.60066, 0.76204] | -1.92 | 2.45-2.72 |
| | | | [0.72090, 0.80097, 0.76056] | | |
| | Conf.8 | | [0.43503, 0.75017, 0.75359] | -1.89 | 2.30-2.34 |
| | | | [0.64842, 0.75017, 0.75568] | | |
| | | | [0.78618, 0.60406, 0.76281] | | |
| | Conf.9 | | [0.51721, 0.82193, 0.75817] | -2.79 | 2.29-2.47 |
| | | | [0.64043, 0.70128, 0.74949] | | |

Next, we examined the stability and dissociation energetics of $Cl_2$ molecules on the Al-terminated AlN(0001) surface. Four different initial configurations are considered: (i) one with the Cl–Cl bond axis parallel to the [0001] direction (Conf. 2) and (ii) three distinct initial configurations with Cl-Cl bond axis perpendicular to [0001] (Confs. 5, 7, and 9 of Table 3). In all cases, the final states are two atomic Cl species adsorbed at the Al-bridge sites, with Cl-Cl separation in the range of 4.69–5.42 Å.

For configuration 9 of Table 3, where the Cl-Cl bond axis is perpendicular to [0001], the increase in the $d_{Cl-Cl}$ of the $Cl_2$ molecule from 2.00 Å to 2.60 Å, lower the total free energy by 3.26 eV/molecule in low surface coverage. In this state, the $d_{Al-Cl}$ is around 2.19–2.60 Å. When the $d_{Cl-Cl}$ extended further to 3.28 Å, the energy decreased by an additional 0.83 eV/molecule, with $d_{Al-Cl}$ of 2.16–2.45 Å. Beyond this point, the energy change is small, reaching a minimum at a $d_{Cl-Cl}$ of 3.96 Å, where the $d_{Al-Cl}$ is around 2.36–2.46 Å. This configuration yields a $Cl_2$ dissociation energy of 4.24 eV/molecule. Similar values of dissociation energies are determined for Confs. 5 and 7, at 3.04 eV/molecule and 4.60 eV/molecule, respectively.

For the Conf. 2 of Table 3, where Cl-Cl bond axis is parallel to [0001] of AlN, the increase in the $d_{Cl-Cl}$ from 2.00 Å to 2.56 Å resulted in an energy release of 0.39 eV/molecule. A further increase of to $d_{Cl-Cl}$ 2.62 Å, a significant amount of energy is released (2.54 eV/molecule) as both Cl atoms are close to the surface with $d_{Al-Cl}$ distances of 2.19–2.40 Å. Further increase in the $d_{Cl-Cl}$ to 4.04 Å, an additional 0.72 eV/molecule is released, giving final $d_{Al-Cl}$ of 2.36–2.47 Å. The dissociation energy for this configuration is 3.65 eV/molecule. Thus, we conclude that the dissociation of $Cl_2$ on Al-terminated (0001) surface of AlN is a barrierless and exothermic reaction.

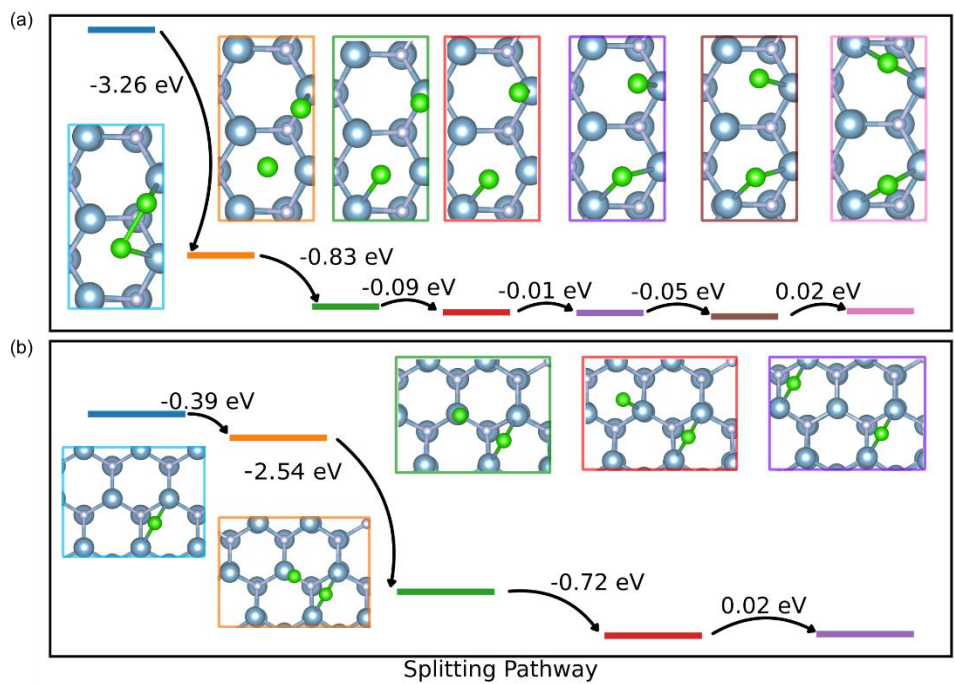

Figure 5: (a) and (b) shows the dissociation pathway of $Cl_2$ molecule with initial configuration of conf.9 and 2, respectively (as mentioned in Table 3). The ball and stick model of each intermediate step is also presented.

Our first-principles calculations revealed that the dissociation of $Cl_2$ on the Al-terminated AlN(0001) surface is energetically favourable. This finding introduces a potentially new paradigm for surface modification in PEALE process. In conventional PEALE processes of group III-nitride, surface modification is typically achieved through exposure to a chlorine containing plasma, which comprises reactive ionic species like $Cl^+$, $Cl_2^+$, $BCl_3^+$, and $BCl_2^+$. The generation and stabilization of such plasmas are inherently time- and energy intensive, requiring sophisticated power delivery units. The present results suggest that the plasma-based ionization of chlorine gas may not be strictly necessary for initiating surface modification on the Al-terminated AlN(0001). Instead, our calculations indicate that the Al-terminated surface itself can act as a catalyst for the dissociation of neutral $Cl_2$ molecules producing chemisorbed atomic Cl species directly on the surface without using plasma ionization. This mechanism could enable a more energy-efficient and cost-effective approach to surface modification in PEALE processes, potentially reducing the plasma power requirements and simplifying the process conditions thereby opening new routes for optimizing etch chemistry, minimizing damage, and enhancing throughput in semiconductor processing.

In summary, we conducted a first-principles DFT simulation to investigate the adsorption, desorption, and diffusion of Cl atoms and molecular $Cl_2$ on Al-terminated (0001) surface of AlN. Our findings indicate that adsorption of both Cl and $Cl_2$ is chemisorption in nature. Among the four symmetric surface adsorption sites examined (Al-atop, Al-bridge, H3, and T4), at low surface coverage the Al-Bridge site is found to be the most energetically favourable for atomic Cl, exhibiting a high adsorption energy of −5.53 eV/atom. CI-NEB calculations revealed a trend in diffusion barrier. The barrier is high when a Cl adatom diffuses from a surface site with a higher coordination number to one with a lower coordination number, and conversely, the barrier is lower when the diffusion occurs from a low-coordination surface site to a high-coordination surface site. The adsorption of $Cl_2$ molecules is compared for configurations where the Cl-Cl bond axis is either parallel or perpendicular to the crystallographic [0001] direction of the AlN surface. The configuration with the perpendicular orientation proved to be more energetically favorable, with an adsorption energy of −2.79 eV/molecule. In contrast, the parallel orientation on the bridge site yielded a significantly lower adsorption energy of −1.30 eV/molecule. Notably, our CI-NEB study revealed that the splitting of $Cl_2$ molecules on the Al-terminated AlN(0001) surface is a barrierless process. This barrierless dissociation of $Cl_2$ on the AlN surface suggests a new and promising method for surface modification. This approach could be faster and more energy efficient than the current state-of-the-art PEALE method.